\documentclass[12pt,epsf,amssymb,qsymbols]{article}
\usepackage{tabularx}
\usepackage{array}
\usepackage{graphics}
\usepackage{graphicx}
\usepackage{epsfig}
\usepackage{amsmath}
\usepackage{amssymb}
\makeatletter

\usepackage{verbatim}

\newcommand{\nn}{\nonumber}
\newcommand{\bea}{\begin{eqnarray}}
\newcommand{\eea}{\end{eqnarray}}
\newcommand{\beq}{\begin{equation}}
\newcommand{\eeq}{\end{equation}}
\newcommand{\gev}{{\rm GeV}}
\newcommand{\mev}{{\rm MeV}}

\newcommand{\pdir}{p\kern -5.2pt\raise 0.2ex\hbox {/}}
\newcommand{\vdir}{v\kern -5.75pt\raise 0.15ex\hbox {/}}
\newcommand{\kdir}{k\kern -5.75pt\raise 0.15ex\hbox {/}}
\newcommand{\epsdir}{\epsilon\kern -5.0pt\raise 0.15ex\hbox {/}}
\newcommand{\bvdir}{\bar{v}\kern -5.75pt\raise 0.15ex\hbox {/}}
\newcommand{\Ddir}{D\kern -7.75pt\raise 0.20ex\hbox {/}}
\newcommand{\ldir}{l\kern -5.0pt\raise 0.2ex\hbox{/}}
\newcommand{\varepsdir}{\varepsilon\kern -5.5pt\raise 0.15ex\hbox{/}}

\def\negcdot{\negmedspace\cdot\negmedspace}
\makeatother

\begin{document}
\begin{flushright}
\begin{tabular}{l}
{\tt LPT Orsay,04-46}
\end{tabular}
\end{flushright}
\begin{center}
\vskip 1.9cm\par
{\par\centering \LARGE \bf On the mass differences between}\\
\vskip 0.258cm\par
{\par\centering \LARGE \bf  the scalar and pseudoscalar}\\
\vskip 0.18cm\par
{\par\centering \LARGE \bf heavy-light mesons}\\
\vskip 0.75cm\par
\scalebox{.9}{\par\centering \large  
\sc Damir~Be\'cirevi\'c$^a$, Svjetlana~Fajfer$^{b,c}$ and Sa\v{s}a~Prelov\v{s}ek$^{b,c}$}
{\par\centering \vskip 0.5 cm\par}
{\sl
$^a$ Laboratoire de Physique Th\'eorique (B\^at 210)~\footnote{Unit\'e mixte de
Recherche du CNRS - UMR 8627.}, Universit\'e Paris Sud, \\
Centre d'Orsay, 91405 Orsay-Cedex, France.\\                                   
\vspace{.25cm}
$^b$ J.~Stefan Institute, Jamova 39, P.O. Box 3000,\\
1001 Ljubljana, Slovenia.\\
\vspace{.25cm}
$^c$
Department of Physics, University of Ljubljana,\\
 Jadranska 19, 1000
Ljubljana,
Slovenia.  }\\
{\vskip 0.25cm \par}
\end{center}

\vskip 0.25cm
\begin{abstract}
We discuss the recent experimental observation which suggested 
that the mass difference between the scalar and pseudoscalar heavy-light 
mesons is larger for the non-strange states than for the strange ones. 
After computing the chiral corrections in the heavy quark limit 
we show that, contrary to experiment, the mass difference in the 
non-strange case should be smaller. 
\end{abstract}
\vskip 0.2cm
\setcounter{page}{1}
\setcounter{footnote}{0}
\setcounter{equation}{0}
\noindent

\renewcommand{\thefootnote}{\arabic{footnote}}

\setcounter{footnote}{0}

\vspace*{10mm}
\section{Conflict between theory and experiment}

High statistics $B$-factory experiments at {\sc BaBar} and Belle, besides providing 
the substantial information about the CP-violation in the processes involving $B$-mesons,
also allowed for a precision measurement of the $D$-meson spectrum. 
Together with CLEO and FOCUS, all four experiments reported the presence of the narrow 
scalar ($J^P=0^+$) and axial ($J^P=1^+$) states~\cite{Ds-experiments}, 
the average of which is found to be
\bea\label{exp}
m_{D_{s0}^\ast}^{(0^+)}=2317.0(4)~\mev\,,\qquad
m_{D_{s1}^\prime}^{(1^+)}=2458.2(1.0)~\mev\,.
\eea 
These results were somewhat surprising because both the scalar and axial 
states are below the threshold of their dominant --Zweig allowed-- modes, namely 
$m_{D_{s0}^\ast} < m_D+m_K$  and $m_{D_{1s}^\prime} < m_{D^\ast}+m_K$. 
Therefore the newly observed states are very narrow, contrary to what has been predicted 
by many potential quark models~\cite{oldmodels}.~\footnote{
Note however that such a low scalar state mass was anticipated in 
the model calculation of ref.~\cite{ad}.} This motivated many authors to either generalise the quark 
model potentials as to accommodate the narrowness of the mentioned states~\cite{scalars_are1}, 
employ the unitarised meson model to the charmed scalar states~\cite{unitarised}, or to 
revive the old ideas about the molecular structure 
of these excitations~\cite{scalars_are2}. Before attributing something exotic to 
the states~(\ref{exp}), one should carefully check if the minimal ``quark-antiquark" picture, 
which has been so successful in the history of hadron spectroscopy, 
indeed fails. Such a test cannot be made by insisting on the accuracy of the quark models 
at a percent level because of the questionable contact of any specific quark model parameter 
with QCD. A reliable test of compatibility between 
the ``quark-antiquark" picture and the observed hadronic spectra could be made 
by means of the fully unquenched high statistics lattice studies, which are unfortunately 
 not yet available. The two partially quenched lattice studies, that appeared after 
the announcement of the experimental numbers~(\ref{exp}), reached  
two different conclusions: while ref.~\cite{bali} suggests that the new states 
are unlikely to be the scalar and axial quark-antiquark mesons, 
in ref.~\cite{ukqcd} the difference between the scalar and pseudoscalar 
charm-strange mesons is shown to be consistent with the experimentally measured 
ones.~\footnote{Compatibility with observation was also claimed on the basis of results obtained 
by using the QCD sum rules~\cite{dai}.}

When cataloging the heavy-light mesons it is customary to use the heavy quark spin 
symmetry according to which the total angular momentum of the light degrees of freedom 
($j_\ell^P$) is a good quantum number so that the heavy-light mesons come in doublets of 
a common $j_\ell^P$, e.g.,
\bea
\underbrace{[D_{(s)}(0^-), D_{(s)}^\ast(1^-)]}_{j_\ell^P=\frac{1}{2}^- (L=0)}\,,\quad
\underbrace{[D_{0(s)}^\ast(0^+), D_{1(s)}^\prime(1^+)]}_{j_\ell^P=\frac{1}{2}^+ (L=1)}\,,\quad
\underbrace{[D_{1(s)}(1^+)], D_{2(s)}^\ast(2^+)]}_{j_\ell^P=\frac{3}{2}^+ (L=1)}\,,\quad \dots
\eea
where the index ``s" helps distinguishing the strange from non-strange heavy-light mesons.

After comparing to  the well known lowest states (belonging to 
$j_\ell^P=\frac{1}{2}^-$)~\cite{pdg}, we see that the splittings
\bea\label{strange}
\Delta m_s(0)\equiv  m_{D_{0 s}^\ast} - m_{D_{s}} = 348.4(9)~\mev\,,\quad
\Delta m_s(1)\equiv  m_{D_{1 s}^\prime} - m_{D_{s}^\ast} = 345.9(1.2)~\mev\,,
\eea
are equal. In other words the hyperfine splitting in the first orbitally excited doublet   
is indistinguishable from the one in the ground state doublet. Although various quark 
models give different numerical estimates for $\Delta m_s(0)$, almost all of them share 
a common feature, namely this orbital splitting remains almost unchanged after replacing the 
light $s$- by $u$- or $d$-quark. The surprise (now for real) 
actually came from experiment when Belle reported~\cite{trabelsi}
\bea\label{non-strange}
\Delta m_u(0)\equiv  m_{D_{0}^\ast} - m_D = 444(36)~\mev\,,\quad
\Delta m_u(1)\equiv  m_{D_{1}^\prime} - m_{D^\ast} = 420(36)~\mev\,,
\eea
clearly larger than the ones with the strange light quark~(\ref{strange}), even though the  
error bars in the non-strange results are much larger which reflects the experimental 
difficulty in identifying the broad states. The confirmation of this phenomenon 
came recently by FOCUS~\cite{focus}, namely, 
\bea
m_{D_{0}^\ast} =2407(21)(35)~\mev\;\Rightarrow\; \Delta m_u(0) = 538(41)~\mev\,.
\eea
This truly surprising phenomenon requires an explanation. Since, to a very good 
approximation, $\Delta m_{u,s}(0)= \Delta m_{u,s}(1)$, we shall concentrate 
on $ \Delta m_{u,s}(0)$ and argue that the experimentally established inequality 
\bea
[ \Delta m_u(0) - \Delta m_{s}(0)]^{\rm exp.} >\ 0 \,,
\eea
is in conflict with theory if the phenomenon is examined by means 
of chiral perturbation theory (ChPT). A similar conclusion has been reached 
by the model calculations of ref.~\cite{thorsten}.

\section{Chiral lagrangian for doublets of heavy-light mesons \label{ratiosRBRP}} 

The lagrangian that is necessary for studying the mass difference between 
the $\frac{1}{2}^+$ and $\frac{1}{2}^-$ heavy-light states is~\cite{casalbuoni}
\bea
&&{\cal L} = {\cal L}_{\frac{1}{2}^-} + {\cal L}_{\frac{1}{2}^+}  + {\cal L}_{\rm mix} + {\cal L}_{\rm ct.}\,, \nn \\
&& {\cal L}_{\frac{1}{2}^-} = i {\rm Tr}\left[ H_b v\negcdot D_{ba} \bar H_a\right]
+  g  {\rm Tr}\left[ H_b \gamma_\mu \gamma_5 {\bf A}^\mu_{ba} \bar H_a\right]\,,\nn\\
&& {\cal L}_{\frac{1}{2}^+} = - {\rm Tr}\left[ S_b(i v\negcdot D_{ba} + \Delta_S)\bar S_a\right]
+  \tilde g  {\rm Tr}\left[ S_b \gamma_\mu \gamma_5
{\bf A}^\mu_{ba} \bar S_a\right]\,,\nn\\
&&{\cal L}_{\rm mix} =  h {\rm Tr}\left[ S_b \gamma_\mu \gamma_5 {\bf A}^\mu_{ba} \bar H_a\right] +
{\rm h.c.}\,,\nn\\
&&{\cal L}_{\rm ct} =  {\rm Tr}\left[\left(\lambda \bar H_a H_b  - \widetilde \lambda\bar S_a S_b\right) \left( \xi  {\cal M}\xi+\xi^\dagger {\cal M}\xi^\dagger \right)_{ba}\right]
\nn\\
&& \hspace*{12mm}+ {\rm Tr}\left[\left(\lambda^\prime \bar H_a H_a  - \widetilde \lambda^\prime\bar S_a S_a\right) \left( \xi  {\cal M}\xi+\xi^\dagger {\cal M}\xi^\dagger \right)_{bb}\right]\,,
\eea 
where the fields of pseudoscalar ($P$), vector ($P^{\ast}_\mu$), scalar ($P_0$) and axial
($P^{\ast}_{1\ \mu}$) mesons are organised in superfields 
\bea
&&H_a(v) = {1 +  \vdir \over 2} \left[ P^{\ast\ a}_\mu (v)\gamma_\mu - P^a (v)\gamma_5\right]\;,\qquad
\overline H_a(v) = \gamma_0 H_a^\dagger (v) \gamma_0\,,\nn \\
&&S_a(v) = {1 +  \vdir \over 2} \left[ P^{\ast\ a}_{1\ \mu} (v)\gamma_\mu \gamma_5 - P_0^a (v)
\right]\;,\qquad
\overline S_a(v) = \gamma_0 S_a^\dagger (v) \gamma_0\,,
\eea
with ``$a$" and ``$b$" labelling the light quark flavour. In addition 
\bea
&&D^\mu_{ba}H_b = \partial^\mu H_a - H_b{\bf V}^\mu_{ba}
=  \partial^\mu H_a -  H_b {1\over 2}[ \xi^\dagger \partial_\mu \xi +
\xi \partial_\mu \xi^\dagger ]_{ba}\;,\nonumber \\
&&{\bf A}_\mu^{ab}
= {i\over 2}[ \xi^\dagger \partial_\mu \xi -
\xi \partial_\mu \xi^\dagger ]_{ab} \;,\nn\\
&&\xi =
\sqrt{\Sigma}\,,\qquad \Sigma = \exp\left(2i\frac{\phi}{f}\right)\,,
\eea
with $f\approx 130$~MeV, ${\cal M}={\rm diag}(m_u,m_d,m_s)$, and $\phi$ the usual matrix of pseudo-Goldstone bosons,  
\bea
\phi =\left(
\begin{array}{ccc}
{1\over \sqrt{2} } \pi^0  +  {1\over \sqrt{6} } \eta  &
 \pi^+  & K^+ \\
 \pi^-    &   -{1\over \sqrt{2} } \pi^0  +  {1\over \sqrt{6} } \eta  & K^0\\
K^-& \bar K^0  &-{2\over \sqrt{6} } \eta  \\
\end{array}
\right)\;.
\eea
$g$ and $\tilde g$ are the couplings of the Goldstone boson to the pair of heavy-light mesons 
with $j_\ell^P=\frac{1}{2}^-$ and $\frac{1}{2}^+$, respectively.~\footnote{The coupling $g$ is
proportional to the commonly used coupling $g_{D^\ast D\pi}$, whereas $\tilde g$ is proportional to 
$g_{D_0^\ast D_1^\prime \pi}$.}  $h$, 
instead, is the coupling of a Goldstone boson and the heavy-light mesons belonging to different 
heavy quark spin doublets, namely one meson is $\frac{1}{2}^-$ and the other $\frac{1}{2}^+$ state. 
The meson masses are $m_{H_{{\frac{1}{2}}^\pm}} = m_Q + {\cal E}_{{\frac{1}{2}}^\pm}$, whereas the difference 
between the binding energies in the first orbital excitation and in the lowest lying heavy meson states is denoted
 by $\Delta_S =  {\cal E}_{{\frac{1}{2}}^+}-{\cal E}_{{\frac{1}{2}}^-}$.

\section{Chiral correction to the mass of heavy-light mesons}

\begin{figure}[h]
\vspace{3mm}
\begin{center}
\begin{tabular}{@{\hspace{-0.25cm}}c}
\epsfxsize16.7cm\epsffile{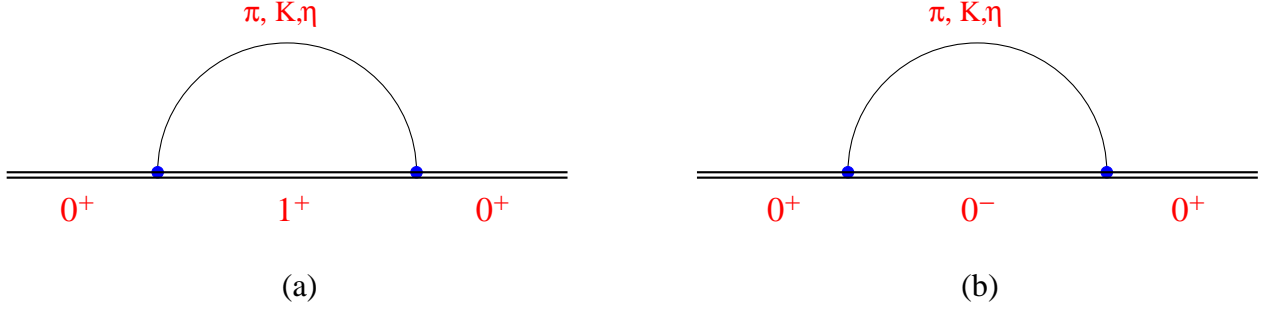}    \\
\end{tabular}
\caption{\label{fig1}{\footnotesize 
Graphs contributing to the chiral shift in the binding energy of the scalar meson $J^P=0^+$. 
By flipping all the parity signs, one gets the graphs relevant to the case of pseudoscalar meson also discussed in
the text. } }
\end{center}
\end{figure}

Since we work in the heavy quark limit, the heavy-light meson propagator 
is a function of $v\negcdot k$ only. $k_\mu = p_{P\mu} - m_Qv_\mu$, is 
the momentum of the light degrees of freedom in the
heavy-light meson. The chiral dressing of the $\frac{1}{2}^-$-meson propagator, 
\bea
G^q_{\frac{1}{2}^-}(v\cdot k) = {i\over 2 v\negcdot k} +{i\over 2 v\negcdot k} 
\bigl(-i\Sigma_q(v\negcdot k) \bigr) {i\over 2 v\negcdot k} +\dots   \,,
\eea 
generates a shift to its binding energy, ${\cal  E}_{\frac{1}{2}^-}\to {\cal  E}_{\frac{1}{2}^-} 
+ \delta {\cal E}^q_{\frac{1}{2}^-}$, where 
\bea
\delta {\cal E}^q_{\frac{1}{2}^-} ={1\over 2} \lim_{v\negcdot k\to 0} \Sigma_q(v\negcdot k)\,. 
\eea
Similarly,
\bea
G^q_{\frac{1}{2}^+}(v\negcdot k) = {i\over 2 \left( v\negcdot k - \Delta_S\right) } +
{i\over 2 \left(v\negcdot  k - \Delta_S\right)} 
\left(-i\widetilde \Sigma_q(v\negcdot k)\right){i\over 2 \left( v\negcdot k - \Delta_S\right) } +\dots   
\eea 
leads to 
\bea
\delta  {\cal E}^q_{\frac{1}{2}^+} ={1\over 2} \lim_{v\negcdot k\to 0} 
\widetilde \Sigma_q(v\negcdot k)\,. 
\eea
Therefore the mass splitting between  $\frac{1}{2}^+$ and $\frac{1}{2}^-$ states in the heavy quark limit is
\bea\label{eq14}
\Delta m_q(0) = \Delta_S + \delta {\cal E}^q_{\frac{1}{2}^+} - \delta
 {\cal E}^q_{\frac{1}{2}^-}\,,
\eea 
where the light valence quark in the heavy-light meson, $q\in \{u/d,s\}$. We will work in the isospin limit,
$m_u=m_d=m_{u/d}$. We focus onto the scalar meson and compute the chiral loop corrections 
illustrated in fig.~\ref{fig1}. 
\bea
-i \widetilde \Sigma^{(a)}_q(v\negcdot k)=\sum_{i=1}^8\sum_{a=1}^3 
\int {d^4p\over (2\pi)^4}\ {-2 \tilde g p^\alpha\over f} \left( t^{i\dagger}\right)_{qa}  {-i \left( g_{\alpha\beta} - 
v_\alpha v_\beta\right)\over 2v\negcdot (k+p)} \ 
{2 \tilde  g p^\beta\over f}\left( t^i\right)_{aq} \ {i\over p^2 - m_{i}^2}\ .
\eea
This integral is standard and the result is expressed in terms of functions $J_{1,2}$ (explicit expressions can be 
found in, for example, 
Appendix~A of ref.~\cite{jsd}) leading to
\bea
&&-i \widetilde \Sigma^{(a)}_q(v\negcdot k)=\sum_{i=1}^8
{-2 \tilde g^2\over f^2} \left(t^i t^i\right)_{qq}  
{3 i\over (4\pi)^2} (v\negcdot k) J_1(m_i,-v\negcdot k)\nn\\
\Rightarrow &&\lim_{v\negcdot k\to 0} 
\widetilde \Sigma_q^{(a)}(v\negcdot k) = - { 6\tilde g^2\over (4 \pi f)^2}
\sum_{i=1}^8 \left(t^i t^i\right)_{qq} {2 \pi\over 3} m_i^3\ ,
\eea
In a completely analogous way, after exchanging ``$+$"$\leftrightarrow$``$-$" in the graph (a), we have
\bea
\lim_{v\negcdot k\to 0} \Sigma_q^{(a)}(v\negcdot k) = - { 6 g^2\over (4 \pi f)^2}
\sum_{i=1}^8\left(t^i t^i\right)_{qq} {2 \pi\over 3} m_i^3\ .
\eea
As for the diagram (b) we obtain
\bea
-i \widetilde \Sigma^{(b)}_q(v\negcdot k)&=&\sum_{i=1}^8\sum_{a=1}^3 
\int {d^4p\over (2\pi)^4}\ {-2 h v_\alpha\over f} \left( t^{i\dagger}\right)_{qa}  
{i p^\alpha p^\beta \over 2[ v\negcdot (k+p) - (-\Delta_S)]} \ 
{2 h v_\beta\over f}\left( t^i\right)_{aq} \ {i\over p^2 - m_{i}^2 }\nn\\
&&\hspace*{-21mm}=-{2 i h^2\over (4\pi f)^2} \sum_{i=1}^8\left(t^i t^i\right)_{qq} (- \Delta_S-v\negcdot k) \left[
J_1(m_i,-\Delta_S-v\negcdot k) + J_2(m_i,-\Delta_S-v\negcdot k)\right]\,,
\eea
and therefore
\bea\label{ref100}
\lim_{v\negcdot k\to 0}  \widetilde \Sigma_q^{(b)}(v\negcdot k) =  -{ 2 h^2 \Delta_S\over (4 \pi f)^2}
\sum_{i=1}^8 \left(t^i t^i\right)_{qq} \left[ J_1(m_i,-\Delta_S) + J_2(m_i,-\Delta_S)\right]\ .
\eea
Similarly,  
\bea
\lim_{v\negcdot k\to 0}  \Sigma_q^{(b)}(v\negcdot k) =  { 2 h^2 \Delta_S\over (4 \pi f)^2}
\sum_{i=1}^8 \left(t^i t^i\right)_{qq} \left[ J_1(m_i,\Delta_S) + J_2(m_i,\Delta_S)\right]\ .
\eea
Notice that compared to eq.~(\ref{ref100}) the sign in front of $\Delta_S$ in the argument of the functions $J_{1,2}$
is now changed. This reflects the fact that the intermediate heavy-light meson, with respect to 
the mass of the meson in the external leg, is now heavier.

After collecting the above expressions into eq.~(\ref{eq14}), we arrive at
\bea
\Delta m_q (0) &=& \Delta_S
 \left( 1 - {  h^2 \over (4 \pi f)^2}  \left(t^i t^i\right)_{qq}\sum_{z=\pm}\left[
 J_1(m_i,z\Delta_S) + J_2(m_i,z\Delta_S)\right]\right)\nn\\
 && + {g^2 - \widetilde g^2\over 8\pi f^2} \left(t^i t^i\right)_{qq} m_i^3 + 
 2 ( \tilde \lambda - \lambda) m_q + 2 ( \tilde \lambda^\prime - \lambda^\prime ) (m_u+m_d+m_s)\,,
\eea
where in the last line we also included the counterterms, thus completing the 
NLO chiral corrections to the mass splitting we consider. The integrals 
$J_{1,2}$ also carry an implicit dependence on the scale $\mu$ which cancels against 
the one in $\tilde \lambda^\prime - \lambda^\prime$. Finally, in evaluating the integrals 
$J_{1,2}$, we set $\bar \Delta =0$ (see eq.~(44) of ref.~\cite{jsd}).

Note that in our loop calculations we include 
the light pseudogoldstone bosons only. The inclusion of light resonances, 
such as $\rho$, $K^\ast$, $\phi$, would involve higher orders in chiral
expansion which is beyond the scope of the approach 
adopted in this paper~\cite{ecker}.

\section{Chiral enhancement or suppression?}

To examine whether or not the apparent chiral enhancement observed experimentally can be 
explained by the approach adopted in this letter we need to consider  
\bea\label{splitt}
\Delta m_{u/d}(0)-\Delta m_s(0) &=& 
{  h^2 \Delta_S\over (4 \pi f)^2} \sum_{z=\pm}\left[ J_1(m_K,z\Delta_S) + \frac{1}{2}J_1(m_\eta,z\Delta_S) -\frac{3}{2}J_1(m_\pi,z\Delta_S)\right.\nn\\
&&\hspace*{13mm}\left.+ J_2(m_K,z\Delta_S)  + \frac{1}{2}J_2(m_\eta,z\Delta_S)-\frac{3}{2} J_2(m_\pi,z\Delta_S) 
\right] \nn\\
&&-{g^2 - \widetilde g^2\over 16\pi f^2}\left(m_\eta^3+2m_K^3-3m_\pi^3\right) -
2 (  \lambda -\tilde \lambda) (m_{u/d}-m_s)\,.
\eea
By using the Gell-Mann formulae,
\bea \label{gell-mann}
m_\pi^2 = 2B_0 m_s  r \;,
\quad 
m_K^2 = 2 B_0 m_s {r + 1\over 2}\;,
\quad m_\eta^2 = 2B_0m_s {r+2\over 3}\;,
\eea
where $r=m_{u/d}/m_s$ and  $2 B_0 m_s = 2 m_K^2-m_\pi^2=0.468\ \gev^2$, we can simply plot the eq.~(\ref{splitt}) 
against the variation of $r$, the light quark mass with respect to the strange quark which is kept fixed to its 
physical value. Before doing so we discuss our choice of values for the couplings $h$, $g$ and $\widetilde g$, and 
for the low energy constant $\lambda-\tilde \lambda$. 
\begin{itemize}

\item $g$-coupling has been  determined  experimentally from the width of the charged $D^\ast$-meson, 
{\underline{$g=0.61(1)(6)$}}~\cite{cleo-g}.~\footnote{A short review of lattice and QCD sum rule estimates of this quantity  
can be found in ref.~\cite{rev-durham}.}

\item There is no experimental determination of the axial coupling in the orbitally excited 
doublet, $\widetilde g$. While the nonrelativistic quark model predicts $\vert \widetilde g/g \vert = 1/3$, a relativistic 
model which correctly predicted $g$ before it was measured~\cite{alain}, one gets $ \widetilde g  = 0.03$. 
The QCD sum rule based estimates are $ \widetilde g = 0.10(2)$~\cite{colangelo}. To cover the whole range of values 
we will take {\underline{$\widetilde g = 0.2(2)$}}.

\item The experimental situation with $h$, the pionic coupling between mesons belonging to different doublets, 
is less clear. If we take the mass and width of the scalar meson as measured by Belle~\cite{trabelsi}, 
we get $h=0.78(9)(8)$, while those measured by FOCUS~\cite{focus} give $h=0.56(8)(6)$, in a very good 
agreement with the QCD sum rule estimates $h=0.60(13)$~\cite{colangelo}. 
From the recent lattice computation of the width of the scalar heavy-light 
state~\cite{michael}, we deduce $h=0.62(6)(4)$, where we 
used the scalar meson mass measured by Belle [$m_{D_0^\ast} = 2308(36)~\mev$], which is more reliable than 
the one measured by FOCUS in that Belle properly separate $0^+$ and $1^+$ 
signals.~\footnote{We thank the referee for drawing our attention to this
point.} The model of
ref.~\cite{alain} predicts $h\simeq 0.54$. To take the 
full spread of the mentioned values we will use {\underline{$h=0.6(2)$}}.

\item In the recent unquenched lattice study~\cite{green}, it has been shown that the splitting that we discuss in this
letter changes very weakly when the light quark is varied between $r=0.65$ and $r\simeq 1$. We will then fix the value of 
$K$ in $2(\lambda -\tilde \lambda)(m_{u/d}-m_s)$ $\to$ $K (m_\pi^2-m_K^2)$, 
by imposing the limit that eq.~(\ref{splitt}) allows for a variation smaller than or equal to $-50$~MeV, 
for $r\in (0.65,1]$. Limiting values are {\underline{$K(1~\gev) \simeq 0.7~\gev^{-1}$}}, for the variation to $-50$~MeV, and 
{\underline{$K(1~\gev) \simeq 1.3~\gev^{-1}$}}, for no variation at all.~\footnote{These values are obtained by
choosing $\mu=1$~GeV. Had we chosen any other $\mu$, the corresponding $K(\mu)$ would be different but the 
resulting $\Delta m_{u/d}(0)-\Delta m_s(0)$ would obviously remain the same.}

\end{itemize}
\begin{figure}
\vspace*{-0.3cm}
\begin{center}
\begin{tabular}{@{\hspace{-0.25cm}}c}
\epsfxsize10.2cm\epsffile{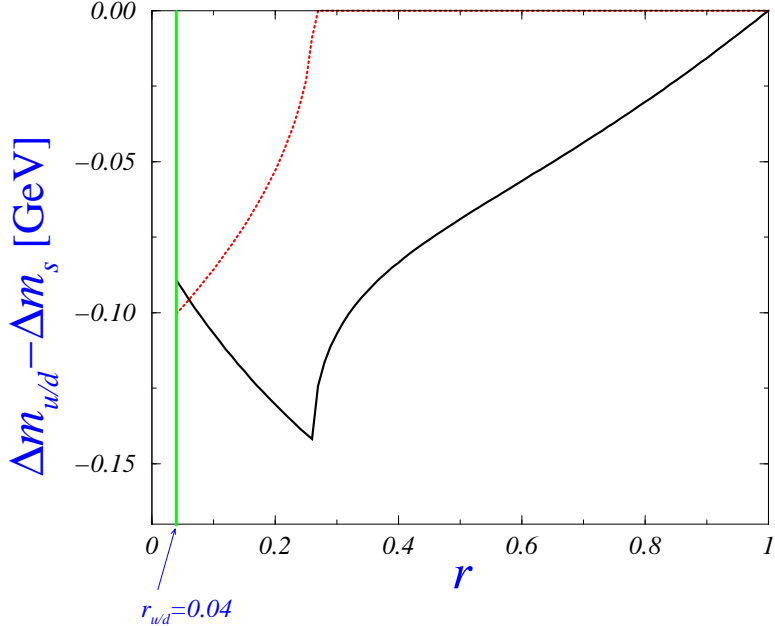}    \\
\end{tabular}
\vspace*{-.1cm}
\caption{\label{fig0}{\footnotesize 
Eq.~(\ref{splitt}) against the variation of $r=m_{u/d}/m_s^{\rm phys}$. The 
physical situations correspond to $r=1$ and $r_{u/d}=0.04$~\cite{leutwyler}.
We used $\Delta_S=0.35$~GeV, the central values for the 
chiral couplings, as discussed in the text, and $K(1~\gev)=0.7~\gev^{-1}$. The imaginary part is depicted 
by the dotted line which reflects the opening of the real pion emission channel
$P_{0}^\ast \to P\pi$.} }
\end{center}
\end{figure}

In fig.~\ref{fig0} we plot the result of eq.~(\ref{splitt}) by using the central values for the couplings 
listed above. In addition we take $\Delta_S= 0.35$~GeV. We see that when the pion becomes lighter than $\Delta_S$, 
the self energies develop the imaginary part, which reflects the fact that the real pion can be emitted via 
$P_0^\ast \to P \pi$. Most importantly, we see that the real part remains always negative 
\bea
\Delta m_{u/d}(0)-\Delta m_s(0) < 0\,,
\eea
contrary to what is experimentally established. 
This conclusion remains as such when varying the parameters in the ranges indicated above. 
The absolute value of the difference of splittings~(\ref{splitt}) depends most strongly on the value 
of the $h$-coupling and it is negative $\forall h \neq 0$. The term proportional to $g^2-\widetilde g^2$ is
negative too. 
It would change the sign only if $\widetilde g^2 > g^2$ which is beyond reasonable doubt. Notice, however,
that it has been argued recently that the doublet of $\frac{1}{2}^+$ states could be the chiral partner of the 
$\frac{1}{2}^-$ doublet, which would imply that $\widetilde g =g$~\cite{eichten}. Even if that assumption was 
indeed verified in nature, our conclusion that eq.~(\ref{splitt}) is always negative, remains true. However,  
as we explained above, from the present theoretical understanding the equality 
between the two couplings, $\widetilde g =g$, does not appear to be plausible.~\footnote{Actually, in any Dirac equation based model, 
$g=\widetilde g$ can be obtained only if one employs the free spinors and set the quark mass to zero.} 

\section{Conclusion}

In this letter we discuss the mass difference of the scalar and pseudoscalar heavy-light mesons. 
Recent experimental observation by Belle and FOCUS suggests that such a difference in the charmed 
mesons is larger for the non-strange light quark than for the strange one, i.e.
\bea\label{c1}
[(m_{D_{0}^\ast} - m_{D})\ - \ (m_{D_{s0}^\ast} - m_{D_s})]^{\rm exp.} >\ 0\,. 
\eea
Such a phenomenon cannot be explained by means of potential quark models in which this difference 
is almost independent of the valence light quark mass. 
We instead used the chiral perturbation theory to examine if the chiral enhancement suggested 
by experiments can indeed be reproduced. After calculating the chiral corrections, we obtain that 
\bea\label{c2}
[(m_{D_{0}^\ast} - m_{D})\ - \ (m_{D_{s0}^\ast} - m_{D_s})]^{\rm theo.} <\ 0\,. 
\eea
This apparent problem remains as such for any reasonable choice of the chiral couplings. 
It should, however, be stressed that our calculation refers to the static 
heavy quark ($m_Q\to \infty$) which might be questionable when discussing the charm quark sector. 
It is nevertheless unlikely that the ${\cal O}(1/m_c^n)$ corrections 
could change the clear qualitative result summarised in eqs.~(\ref{c1},\ref{c2}).

Our observations show that the scalar states are indeed peculiar.
It is probable that the ``quark-antiquark" picture is not adequate
in case of which  the unitarised  meson model of ref.~\cite{unitarised}
or the 4-quark picture for the scalar mesons~\cite{cheng},
which enjoyed success in explaining the spectrum
of light mesons, may be useful remedy in explaining the scalar states
containing one heavy quark.
Further experimental tests, that might prove useful in getting a more definite answer concerning 
the nature of the observed scalar states, were already proposed in ref.~\cite{godfrey}.
  
 
\section*{Acknowledgement} 
It is a pleasure to thank S.~Descotes and A.~Le Yaouanc for discussions 
and comments. The work of S.F. and  S.P. was supported in part by 
the Ministry of Education, Science and Sport of the Republic of Slovenia.


\end{document}